**Examination of the claim by Loram and Tallon that the energy-resolved STM results, in their apparent inhomogeneity, misrepresent the true bulk behaviour of the HTSC cuprates.**

John A. Wilson

H.H. Wills Physics Laboratory

University of Bristol

Bristol BS8 1TL.  U.K.

**Abstract**

An attempt is made at reconciling the results of the prime experiments on the high temperature superconducting cuprates from Loram *et al.* and Davis *et al.* relating to electronic specific heat and to energy-resolved scanning tunnelling microscopy (STM), respectively, over the question of electronic inhomogeneity.  I see the latter, *at the appropriate level*, as being essential to the evolution of HTSC in these materials, and try to bring into alignment the above key works, not only with each other, but with a wealth of related work.  This is undertaken around a negative-*U* scenario long advocated by the author.

Loram and Tallon very recently have restated the claim, looking at the matter from the perspective of their refined electronic specific heat analysis [1], that the energy-resolved STM results from the HTSC cuprates, as presented by J.C.Davis and coworkers [2] and several other groups, would look to relate to surface effects. Their claim is that these results, the famous 'gap maps', cannot reflect what is proceeding as regards HTSC, once away from the surface and sensed by a thermodynamic probe like the specific heat measurement. The relative sharpness of the electronic specific heat peak seen across $T_c$, even in the strongly 2D environment of Bi-2212, Tl-2201, and Hg-1201, and accounting for the strongly fluctuational conditions about $T_c$ (which signal in $C_v(T)$ as in $\rho(T)$ data [3] the presence of preformed pairs) cannot, it is the claim of Loram and Tallon, be at all reconciled with the very significant spread in bulk 'gapping' very often imputed from the STM work. Moreover each gap map, because it is frozen in time, appears not to relate either (directly at least) to the fluctuating stripe phenomenon sensed in bulk X-ray and neutron diffraction work [4].

While there exist bulk sensitive measurements which would support a different view to the one expressed by Loram and Tallon in regard to the degree and form of inhomogeneity active in the HTSC cuprates, these Cu and O NQR and NMR results usually come from LSCO [5], a material displaying special structural complications. The yttrium site NMR from YBCO appealed to now, as earlier, by Loram *et al.* [6] would nonetheless still support a not insignificant degree of local inhomogeneity, even at the static level. µSR [7], unlike NMR, is drawn specifically to sample, on a faster time scale, around negatively charged 'imperfections', and in the HTSC cuprates these are legion, being deliberately inserted to transform the $CuO_2$ layers there to metallicity and superconductivity. It was µSR work which in fact first endorsed the view I had earlier illustrated in fig.4 of ref. [8] that this highly local inhomogeneity inevitably would play a key role in the character of the superconductivity of these 'doped' Mott insulators, especially in their underdoped regime. Use of the 'Swiss cheese' terminology as *à propros* to the HTSC condition first actually appeared in the µSR work of Nachumi, Uemura, *et al.* [9] dealing with the low level substitution into the cuprates of isoelectronic but energetically dissimilar Ni and Zn for Cu. The doping of Sr for La in LSCO, the extra oxygen in LCO, YBCO, HBCO or BSCCO, the non-isoelectronic counter-substitution of Tl by Cu in TBCO, all of course introduce *charged* lattice defects. Ultimately these become frozen in, and randomly so when samples are rapidly cooled from their high temperature annealing conditions. Where exactly the charge appertaining to these 'dopant' centres subsequently organizes itself is a somewhat different matter. Apart from at $p = {}^1\!/_8$, the charge self-organization is clearly dynamic, not static – see refs. [10] for my understanding of 'stripe' formation and of the role played by the Jahn-Teller effect in this dynamic charge organization and the detailed development of HTSC. Where the Madelung potential most strongly is perturbed by the random doping, there it is possible for the charge stripes to become permanently pinned. It is in fact at certain such stochastically favoured positions that I envisage electron pairing as being most strongly effected. This is because I have taken it from the beginning [11,8] that in HTSC one is encountering a negative-$U$ phenomenon based upon the fluctuational conversion of the state approximating to $^8Cu_{III}^{0}$ over into its $^{10}Cu_{III}^{2-}$ double-loading condition. This is admissible by virtue of the latter's $p^6d^{10}$ shell-closure form which entails the termination of all $p/d$ bonding-antibonding interaction between the Cu and O atoms within the pair-recipient coordination unit. There, the ensuing rearrangement of the now full $p$ and $d$ states, specifically their energy inversion, together with the accompanying local lattice relaxation, procures stability for the $^{10}Cu_{III}^{2-}$ condition, at least on a fluctuational basis. I believe this negative-$U$ state to have been sensed in pump-probe and other appropriate optical work [12], as argued at length in [13]. Pairing obtained at the negative-$U$ centres proceeds in the resonant view embraced [14] (wherein the above double-loading state stands near-degenerate with the chemical potential) to induce pairing over most of the Fermi surface and with a single $T_c$ value. The negative-$U$ centres, despite being inhomogeneous in location and to a certain degree energy, cooperate to bring about a condensed state that is global and phase-locked through the intimate intersite tunnelling of the correlated pairs. In standard superconductivity the pairing envisaged is of retardedly coupled +**k**/-**k** quasiparticles. In locally instigated negative-$U$ pairing the pairing is direct, the pair potential is no longer the classic delta function of the BCS formalism [15], and the coherence length, $\xi$, of the resulting superconductivity is, in keeping with the high $T_c$, high $\Delta$ and high $H_{c2}$, extraordinarily small. In optimally doped HTSC material the value of $\xi^{\perp c}$, as assessed from $H_{c2}^{//c}$, is only ~ 15Å, or about half the inter-stripe spacing [16], with $\xi^{//c}$ even less. The superconductivity emerging in the $p$-type cuprates is deemed essentially $d$-wave ($d_{x^2-y^2}$) in form [17], the maximum gapping being at the saddles of the Fermi surface (i.e. in the basal axial directions of the crystal structure - the Cu-O bond directions there), with the nodes oriented in the 45° directions [18]. The gapping clearly is of strong-coupling form, the observed ratio $2\Delta(T=0)/kT_c$ being *ca*. 5.5 for all the various HTSC cuprate families over a wide range of doping about optimal [19]. Such a value consistently is extractable from $C_v$, NMR, Raman, tunnelling, neutron and ARPES data, and is to be compared with 4.3 for weak-coupling mean-field behaviour within a $d$-wave symmetry setting.

It is very apparent from $C_v$ [20], as from other data, that much higher ratios than 5.5 are to be gained from underdoped material if one persists in relating the 'outer' gapping in evidence there to the value of the superconducting onset temperature, $T_c(p)$ [21]. The problem is that DOS pseudogapping in underdoped material in actuality develops from more than one source upon retreating back towards the Mott-insulating condition of $d_{x^2-y^2}$ state half filling. There comes that associated with RVB spin coupling [22,10c], as well as the residual Mott charge gapping emanating from half filling,. Then too there occurs a pseudogapping local to the negative-$U$ centres when- or wherever these prove insufficient in number or of an energy too far removed from resonant alignment with the chemical potential to support the superconductivity. It is to be recalled that, beyond the enhanced bulk superconducting fluctuational behaviour exhibited by all these materials [23,3], there emerges in addition clear evidence for the presence of local pairs to tens of degrees above $T_c$, on engaging appropriately sensitive probes such as the Nernst effect [24] or tracking the electrical noise [25]. Even very precise measurements of the basal lattice parameters detect incipient susceptibility to the superconducting condition from far above $T_c$ [26].

All this multi-sourced pre-superconductive activity and gapping help to create a pronounced minimum in effective density of states in the centre of a band that if uncorrelated would sit close to a saddle-point maximum. The outcome is that right up to $p$ = 0.3 the Seebeck coefficient is positive at low temperatures due to a negative sign to $(d\sigma/dE)_{EF}$ persisting through to such doping [27]. Similarly the strong temperature dependence of the Hall coefficient [28] arrives because of the changing balance within the $d_{x^2-y^2}$ set of electrons between coherent $p$-type band quasiparticles and non-band-like negative carriers (about the nodes and saddles respectively). The latter become more relevant at very high temperatures, and again at low temperatures upon approaching superconductivity. Throughout, severe and very anisotropic scattering is experienced all around the Fermi surface [29]. At high temperatures this becomes sufficient to bring incoherence and even weak localization to the carriers of the saddles [30], whilst at low temperature these same carriers begin to become abstracted into local bosonic pairs. The saddles provide the scattering 'hot spots' for negative-$U$ pair formation [13b] (just as for potential CDW/SDW nesting [31]). It is most noteworthy that across the underdoped regime, $n_s$, the number of superconducting pairs (assessed from μSR [7,9] and other penetration depth [32] measurements) relates more closely not to $n$, the total number of electrons present in the $d_{x^2-y^2}$ band, but to $p$, the 'doping' content there below half filling. The local character to events remains pre-eminent here. The electrons instigating the HTSC phenomenon are not particularly 'good' quasiparticles, and nor are those throughout the saddle region which primarily respond. This extensive impairment to standard Fermi liquid behaviour is very much in evidence in the ARPES normal state spectra, after one has left the diagonal 'nodal' orientations on the Fermi surface [33].

Correspondingly in the superconducting state one must be very careful in one's reading, therefore, both of the ARPES and the energy-resolved STM spectra, especially where these relate to saddle point states and maximal gapping. One does not have a simple, classical, mean-field, BCS-type situation here, any more than in the normal state one had a classical Fermi liquid. What is remarkable is that the classical approach of the BCS, Eliashberg-Nambu and McMillan-Dynes formalisms can be extrapolated as far into the ultra-correlated regime as looks the case and yet reach a meaningful parametrization [34]. The Uemura plot [35], namely $\log T_c$ vs $\log(n_s^{2/3})$, would portray the superconductors $NbSe_2$, $MgB_2$, $PbMo_6S_8$ and $K_3C_{60}$ as bridge materials between the classical and the cuprate materials, but still they appear reasonably well dealt with by the traditional approach, in spite of their steadily reducing coherence lengths. This is very comparable to what is found within the normal state, where effectively classical quasi-particle Fermi liquid behaviour extrapolates far farther into the highly correlated regime than might have been anticipated – witness the degree of preservation of the classical Lorenz number, $L_o$, for $\kappa_{th}/\sigma T$ [36]. Indeed such a value is approached in very heavily overdoped $(La_{1.7}Sr_{0.3})CuO_4$, even though the prefactor $A$ displayed on the $T^2$ resisistivity behaviour stands five times larger than the Kadowaki-Woods plot would support [37].

One very striking feature with the cuprates, and especially underdoped cuprates, is that the close approach to the onset to superconductivity sees considerable apparent improvement to the quasiparticle mean free path (e.g. the empirical Lorenz number drifts down towards $L_o$ [38]). From thermal Hall data, Ong *et al.* [39] deduced quasiparticle mean free paths growing rapidly towards 10 nm in the run in to $T_c$, this despite the rapid rise in electrical noise which the proximity to stable pairing and bosonization of the active electronic system witnesses [25]. What the effect here might be from preformed pairs is still to be resolved.

The question now arises with both the low temperature ARPES and tunnelling results as to how one is to read off the superconducting gap from these spectra when there is occurring in the underdoped and even optimally doped HTSC materials so much correlated activity beyond that specifically appertaining to the global superconducting state, with its set $T_c$. Indeed Loram and coworkers have themselves earlier expressed earlier their $C_v$ and nmr results in terms of a combination of superconducting gap and pseudogap [40]. As indicated above the former gap is related to $T_c$ by the equation $2\Delta(p,T=0)/kT_c(p) \approx 5.5$. In the underdoped regime $T_c(p)$ drops away from $T_c^{max}$ quadratically as $(1 - (p_o-p)^2)$ to vanish when $p \sim 0.05$ ($p_{o(pt)}$ here $\approx 0.16$) [41]. But it is found as well that into the underdoped regime $n_s$ reduces steadily from optimal doping roughly as $(p_o - p)$, accordingly making $n_s$ not proportional to $T_c$. Rather the remarkable Uemura relation $n_s^{2/3} \propto \log T_c$ holds [35], implicating a Bose-related behaviour. Just as significantly the superconducting condensation energy, $E_c$, extracted from $C_v$ is proportional neither to $n_s$ nor to $T_c$. Indeed it emerges as being a steeply augmenting function of $p$, coming to a sharply cusped maximum at $p = 0.185$, somewhat beyond where $T_c$ smoothly maximizes [40]. Note that where $T_c$ maximizes at $p = 0.155$, $E_c$ still is only one third of what it is to become by $p = 0.185$. Actually $p = 0.185$ is very close to the $p$ value at which the specific heat work and related assessment consistently indicate the pseudogap to be rapidly vanishing [42]. It is, perhaps, this doping level which marks the high point to HTSC behaviour as electrons are brought most effectively into the superconducting condition. For $p$ beyond this concentration, despite the value of $n_s$ continuing (for a short while at least) to rise, already both $T_c$ and the condensation energy per mole are well in decline. Niedermayer *et al* [43] in fact report for Tl-2201 that $T_c$, $n_s$ and $p$ are interrelated across the full range of $p$ by the expression

$$n_s/p \propto m^*.[1 + \sqrt{p}.(T_c(p_o)/T_c(p))] \quad .$$

This confers upon the $T_c$ versus $n_s$ plot (with $p$ as running parameter) a characteristic, inclined and sharply lobed shape, often referred to as the 'butterfly wing' or 'boomerang effect'. Note that in YBCO$_{7-\delta}$ as $\delta \rightarrow 0$ and virtually all the chain electrons there finally are carried through to superconductivity, this incorporation occurs without any growth in $T_c$ [44]. I have in [10a] taken the above planar 'hole' concentration of $p = 0.185$ to afford the optimized combination of doubly-occupied negative-$U$ centre population and energy location. Namely a sizeable population coming into resonance at the hot spot with the chemical potential, at a level of metallicity that is high enough to uphold marginal fermi liquid behaviour, and yet not so high as to overly screen the local Madelung potential requisite for local pair creation at the negative-$U$ centres.

Regrettably most of the HTSC community has no regard for negative-$U$ centres. The physics of HTSC is seen solely as the preserve of $d_{x^2-y^2}$ or $T_{1g}$ symmetry [17]. Where quasiparticles of a more standard form exist then acquires $T_{2g}$ 'nodal' symmetry [45]. There has developed an insistent emphasis upon aspects of the physics holding these two symmetries, a quasi-standard fermionic physics, as, for example, in the purported SDW formation to issue from the saddle 'hot spots' of the Fermi surface [46]. A further case is to be found with the detailed review of the electronic Raman results very recently released by Devereaux and Hackl [47], where all emphasis is placed on the $T_{1g}$ and $T_{2g}$ anomalous spectra to the detriment of the somewhat more difficult to acquire $A_{1g}$ spectra. The latter spectra, however, are even more striking, but are confined in [47] as a 'mystery' to the comments in the conclusion. It is, note, the $A_{1g}$ peak which at 330 cm$^{-1}$ (41 meV) in $T_c^{max}$ ~ 95 K systems would correspond to $2\Delta_0$ ~ 5½$kT_c$. What is more, it is this $A_{1g}$ gapping which exhibits the temperature and doping dependence [48] anticipated of global superconductivity – indeed the same $2\Delta$ behaviour as tracks the much examined $(\pi,\pi)$ 'resonance' peak found below $T_c$ in inelastic neutron work [49], the spin-flip pair excitation [50]. (The latter feature, note, pointedly displays no isotope effect). The $T_{1g}$ and $T_{2g}$ electronic Raman peaks are in contrast not reached until 440 cm$^{-1}$ (55meV) or more, up in the range of the pseudogap phenomena [51]. In comparable fashion standard infrared work casts its emphasis upon an even higher energy pseudogap region, strong IR gapping developing below around 800 cm$^{-1}$ or 0.1 eV (i.e. ~$J$) and coming to dominate the optical self-energy, so helping to mask any action at $2\Delta$ directly relating to the superconductivity itself [52]. Cardona was the first to stress the key behaviour of the $A_{1g}$ electronic Raman spectrum [53] and to show that the peak energy arises exactly where small sharp discontinuities in the phononic Raman spectrum mark the 4 K superconductive gap as such to reside [54]. $A_{1g}$ symmetry covers not only isotropic $s$-wave superconducting symmetry but, in general, extended $s$-wave symmetry as well. Nodes in the latter can very suitably assist at the Cu coordination unit centre in holding down the +$U$ repulsive component to the overall $U$ value.

How now then are we to read the underdoped gap 'bit maps' produced by Davis *et al.* from their energy-resolved STM experimentation [2]? If not as extremely as Davis *et al* seem inclined, then certainly not either as Loram and Tallon would construe them [1]. They imply that if the pseudogap energy were to be significantly inhomogeneous, the superconducting gap necessarily is going to be inhomogeneous too. And, if indeed inhomogeneity were to exist on anything approaching the scale of the mapping, then the superconducting onset temperature ought itself to be much more disordered than is patently the case from the electronic specific heat peak behaviour recorded across the superconductive onset. This view rests, however, upon treating the superconductivity as being of rather standard form, fermion based and density of states driven. All change in the local condition in such a case would be directly reflected in local $T_c$ values, particularly if, as the bit maps are taken to imply, the superconductive gap value is at all local. Yet $T_c$ from the $C_v$ results, as from other data, does not mirror any such spread. Therefore it is claimed in [1] that the inhomogeneity evident in the STM map must in its entirety be a surface effect and cannot in any way extend to the circumstances of the bulk.

However, if one looks carefully at the 4 K STM results, as indeed was originally pointed to by McElroy *et al* themselves in [2a], the inner part of the gap below $2\Delta \sim 30$ meV actually *is* rather invariant across the field of view. It is only the outer wings of the STM spectrum which stand strongly inhomogeneous – and it is the latter that dominate the bit map as customarily it is generated (via use of an overall $(dI/dV)^{max}$). Already though we have seen these increased energies relate to the pseudogap, not to the superconductivity itself. Close in to $T_c$ the thermodynamics is governed by the opening up of the deeper inner gap, without discontinuous change there to the pseudogap. The coherence length of the superconductivity remains in excess always of the dopant nearest neighbour separation at the levels of substitution involved in HTSC samples (viz. $p > 1/18$, giving a typical n.n. spacing of very rarely more than $3a_o$). Within the random distribution of dopants the movement of quasiparticles is not prevented once above $p = 0.05$, and even less so will be the transfer of bosonic pairs under microscopic quantum tunnelling. A uniform phase angle duly becomes established right across the field and a unique $T_c$ value results.

What is more problematic for the system, however, is what proceeds in the single particle spin system. The field of the spin pseudogap certainly is disjoint under the variety of $J$ values between sites. While RVB holds in some microregions, in others spin glass behaviour shows up, particularly as one drops the $p$ value and shifts substantially towards the antiferromagnetic condition of the Mott insulator [55]. Ultimately local charge gapping becomes dominant as the $+U$ circumstances of band half-filling and non-metallicity start to prevail. Hence the gap bit maps, as they are customarily generated and presented, automatically are transferring emphasis onto the non-superconductive reaches of this multi-stranded pseudogap, and away from physics directed specifically towards events at $T_c$; these latter are much more homogeneously expressed by the inner gap [2a]. Right at the surface, furthermore, it is inevitable pseudogapping will be favoured by local escape of the dopant oxygen from the opened cleavage plane. Remember that $O^{2-}$ is not a stable species when not fully embedded, and surface $O^{2-}$ is open therefore to restructuring.

A comparable complexity of behaviour is enfolded in the ARPES results, so well publicized but yet again it would now seem so unfortunately interpreted [56]. The weakly defined spectral location of $E_F$ at the saddles in the normal state condition becomes replaced in the superconducting state of Bi-2212 by a contrasting very strong peaking. This peak feature, seen also in other bi- and tri- layer HTSC cuprates, has, despite persisting to 30 K above $T_c$ and with its energy throughout almost temperature independent, become widely ascribed in its entirety to a superconducting 'coherence peak'. The outcome has been that the latter invariably becomes permitted to deform the outer shape of the superconducting gap leading many ARPES papers to cite $2\Delta$ values considerably in excess of the true superconducting gap. This, for example, includes the paper from Mesot *et al* [57] seeking to enumerate the $2\Delta(\theta)$ values around the Fermi

surface – and indeed finding roughly $|d_{x^2-y^2}|$ azimuthal form. The STM papers from the Davis group embrace these same overstated values for $2\Delta$, a matter I have already taken issue with in [14].

As is to be seen in fig.13 from the recent ARPES paper by Peets *et al* [48] it is, however, the leading-edge maximum-gradient (LEM) point in such spectra that properly identifies the real $2\Delta(p,T=0)/kT_c(p)$ ratio ~ 5.5 near $p_o$. The higher energy ARPES peak value actually does not even extrapolate to zero as $T_c^{opt}$ moves to zero under a change of system, unlike the LEM point. Nonetheless the photoemission peak binding energy plainly is diminishing as $T_c^{opt}$ falls, and thus it is not registering a transfer out into the pseudogap in the underdoped regime in the way the STM peak does. The separation between the ARPES peak and LEM energy is discovered to fall steadily between HTSC systems from 16 meV (or 128 cm$^{-1}$) when $T_c^{opt}$ ~ 95 K to 8 meV (or 64 cm$^{-1}$) as $T_c^{opt} \to 0$ K [48]. This separation quite conceivably could relate to the reflectivity edge in the *c*-axis far infra-red spectra identified with the interplanar Josephson plasma resonance and its associated energy loss function [58]. In [58b] the peak in the energy loss for LSCO $p$=0.17 is found at 80 cm$^{-1}$. The ARPES peak would then betoken the latter corporate excitation of the condensed electrons being strongly coupled into the photoemission process. Accordingly the ARPES peak must be viewed, as with the STM peak, as being far removed from the classical coherence peak of a standard superconductor. The above would account quite naturally for why the photoemission peaking is so prominent at the highly correlated saddles, just where by every measure, ARPES included, one is finding the normal state coherence being relinquished under the intense scattering active there – in my own view from local pair formation, boson-fermion interaction, and pair destruction [14b]. In keeping with the above perception it is to be noted that no peak is observed below $T_c$ to decorate the $(0,\pi)$ edge in the corresponding ARPES spectrum from *single*-Cu-layer materials such as Bi-2201 [59] and Hg-1201 [60], where it is much more difficult to secure *c*-axis charge transfer. As a pertinent reminder of just how highly correlated these cuprates are, ARPES brings us now the 'waterfall effect' of complete one-electron *d*-band degeneration once well at a distance from $E_F$ [61].

The above puts us in a position to be able to understand the very recent energy-resolved STM results from J.-H. Lee, J.C. Davis and coworkers [62]. These from the present author's viewpoint are being completely misinterpreted and are very likely to provide a seat of much future confusion, given that these colour maps are so compelling. As previously this new paper takes the peak value to designate the superconducting gap. It then proceeds to measure the coupled bosonic mode energy, $U$, from this peak (as opposed to the LEM point) across to the gradient inflection point on the hump (as against to the dip)[†1]. Accordingly, whilst the value quoted for the coupled modal energy, $U$, actually is roughly correct, that for $2\Delta$ is much overstated. Near optimal doping, values of 60 meV and 40 meV, respectively, are more acceptable [14b]. As is emphasized in [62], the value of $U$ in a given sample varies relatively little across the scanning field of view (typically 50x50 nm), whilst the peak energy here alters by a factor of almost three. This is appropriate after those quantities become made to refer respectively to the global superconductivity and to the broadening 'ancillary' peak feature. The authors of [62] would advance however with fig.5c that $\Delta$ and $U$ (as there being located) are *actively* 'anticorrelated'. In reality what is happening is that the peak feature rides across the dip, diminishing $U$ thereby as it is being extracted in [62]. Naturally $\Delta(\mathbf{r})$ and O($\mathbf{r}$) (the location of the extra oxygen on the surface) are found to be positively correlated (with $\Delta(\mathbf{r})$ made to refer to the broadening peak feature). Finally $U(\mathbf{r})$ and O($\mathbf{r}$) exhibit no correlation, $U(\mathbf{r})$ being effectively global and O($\mathbf{r}$) random. Note $U(\mathbf{r})$ in its shown lack of dependence upon position expressly does *not* manifest the characteristics of a local phonon mode, yet ref.[62], for want of any better perceived option, tries to impute a phononic ('lattice') origin to the above finding. In the scheme I have outlined previously, the above "$U$" at 60 meV relates to the 'Anderson mode' for the local pair condensate, at a slightly deeper binding energy than the more dispersed mode from local pairs existing outside the condensate. It is the latter mode that features in fig. 2 of [14b], as plotted out using the earlier Fourier transformed STM scattering data from Hoffman *et al* [63] and McElroy *et al* [2a].

---

[1] This is a most unfortunate choice of letter, which we write here in cursive form to distinguish it from the Hubbard-*U* energy – positive and negative.

Let us now return to the matter of symmetry, for that must fundamentally reflect what is proceeding in HTSC cuprate physics. ARPES, being involved with $\psi^2$ rather than $\psi$, does not sense the phase angle component to the superconducting order parameter, only its magnitude. With the nodal arrangement for $|d_{x^2-y^2}|$ and extended-$s$ states co-aligned in the 'diagonal' 45° directions (i.e. bisecting the Cu-O basal bond directions), the two channels should cross-couple whenever they relate to $\psi^2$ rather than to $\psi$. But surely that is the circumstance within the current charge-based, negative-$U$ scenario, the latter constrained by the coordination unit's geometry and the tight-binding electron bands. From a phonon normal mode analysis for the cuprate structures, $A_{1g}$ (or $\Gamma_1$) symmetry covers there a coordination unit's longitudinal breathing mode – exactly, one notes, as the electronic Raman data echoes. The charge flow and the lattice respond with the same natural symmetry as modification is incurred to the electron loading of a coordination unit's $d_{x^2-y^2}$ $\sigma^*$-state. Electron energy loss (EELS) experiments [64] indeed put in evidence a very lossy peak with precisely the above indicated 60 meV location. Although this EELS feature appears at $T_c$, its subsequent intensity temperature dependence would imply that the true onset to activity actually is not $T_c$ itself, but for BSCCO-2212 some 30 K above $T_c$, very much as intimated by the Nernst data [24,65].

What additionally commences well above $T_c$ and adheres to this same $\Gamma_1$ or $A_{1g}$ symmetry is found to be an extremely strong and highly anomalous basal softening of the uppermost (80 meV at the zone centre) LO phonon mode, this advancing sharply from around half-way to the zone boundary [66]. This lattice breathing mode is, it would look, coupling very strongly with some short wavelength mode. Now, long ago a zone edge mode seemingly superfluous to the customary phonon normal mode analysis was detected in neutron scattering work – and in YBCO$_7$ it lay around 55 meV (or 440 cm$^{-1}$) [67]. There was observed to arise a striking gain in the line-width/coupling-strength for this A$_{1g}$ symmetry scattering mode in cooling across $T_c$. What is more, upon examining the Raman spectrum over the same energy range one clearly observes developing some underlying, highly temperature dependent activity. Puchkov [68] was the first to emphasize that the Raman activity in this range is of both A$_{1g}$ and B$_{1g}$ symmetry, is expressly coupled to the superconductivity, and is manifesting the above 440/330 or 4/3 energy relation to 2$\Delta$. The upper feature (which can be resonantly enhanced) is noted to adjust its own energy between different HTSC systems in step with $T_c$. Normally in simple phonon-related Raman scattering, because the probe is a photonic one, one is looking there at what transpires near the zone centre. However in electronic Raman work direct excitation can arise from at whatever $k$ states reside 'addressable' fermions, and such should be the case too for the present composite bosons. The bosonic excitation in question here is to be understood [14b,69] as being not that of the Γ point condensate itself, but, as was mentioned above, of the addressable population of free pairs persisting outside the condensate. In the HTSC cuprates the latter mode is centred away from **K**=0 around each saddle point, **K** = (0,π), *etc.*. de Llano *et al* [70] in their theoretical treatment of BCS/BEC crossover have indicated that this boson excitation mode of finite-K pairs should display linear dispersion in K, the local pair momentum, and not the customary quadratic variation. As is presented in [14b], and already alluded to, I regard this mode to constitute the weakly dispersed bosonic mode met with in the Fourier transform STM work [63,2]. GHz spectroscopy provided the first direct evidence of these bosons existing outside the condensate [71]. The above mode has to be distinguished from that relating to the local pair condensate itself and figuring in the ARPES work. The latter very weakly dispersed mode naturally takes the slightly greater binding energy of about 60 meV: the mode for which such a wide variety of origins have been proposed other than that advocated here, all origins invariably of T$_{1g}$ rather than A$_{1g}$ lineage [72]. The upper dispersed pair mode is seen only for $k$-values where it runs between $E_F$ and 2$\Delta^{max}$ [14b]. By contrast, as Belkhir and Randeria have shown [73], the lower condensate mode can form only for $k$ values larger than $\xi^{-1}$ – namely here, where $\xi$ ~4$a_o$, beyond about halfway across the zone – as is observed.

Many BCS/BEC crossover papers do not accommodate any negative-$U$ aspect to events [74] (and the converse likewise is true [75]). It is very apparent, however, that to make decisive headway theoretically with the cuprates will demand that the inhomogeneous nature of these near-localized, mixed-valent systems be appropriately embraced throughout. The stance taken by Loram and Tallon, for example, that the real situation is not in any way as fragmented as the STM work would suggest appears to the present author a highly retrograde one, clearly in conflict with chemical understanding as to where the HTSC systems stand. Their earlier reported observations regarding the effects of a magnetic field upon specific heat data (see [76]) in fact serve to endorse what presently is being baulked at – see section E17 (p.1021) in [10c]. The above comments are just as applicable regarding most magnetic modelling of what occurs in the HTSC cuprates [77].

Where the confusion is arising is in viewing the STM peak as marking the local superconducting gap energy, as if it were the 'coherence peak' of an extrinsically perturbed classical superconductor, when in fact it marks the the negative-$U$ state energy in a highly local, intrinsicially inhomogeneous, non-BCS superconductor: a superconductor at the BCS/BEC crossover and driven by the resonant energy location of an appropriate subset of the negative-$U$ centres. At a given doping many nanometre regions find themselves holding negative-$U$ centres for which the energy is sub-resonant. When an optimally doped sample is replaced by an underdoped less metallic one, the number of subresonant centres climbs as their typical energy falls away from resonance, under the now less strongly screened trivalent Madelung potential to which they owe their existence as metastable doubly-loaded entitites. We are, then, talking about intrinsically different sorts of Swiss cheese, and with the cuprates it is not of the type encountered in $(Nb/Mo)Se_2$ or even $(Mg/Li)B_2$. One saw from the old STM work of Maggio-Aprile *et al* [78] that within field vortices established through an underdoped HTSC sample, the peak section of the tunnelling response persists, whilst it is the inner superconducting part of the signal which becomes suppressed.

This finally brings us to consider the intrinsic level of inhomogeneity of the inhomogeneity itself. That which is being picked up by Davis and colleagues [2,62], being of the scale of 4nm, definitely implies a degree of chemical dopant clustering there that is not just statistical. In the latter case one would expect somewhat less than half this length scale. Is the above situation an outcome of the sample annealing routine, or is it a surface related problem, as pointed to earlier? The fact that Fischer and coworkers [79] since have recorded considerably less striking, presumably 'finer-grained', STM mappings would strongly support the former case. The secondary level of graining being now reported by Lee, Davis *et al* in fig.2 of [62] is much more on the scale of what to expect, and the fact that it displays too some indication of striping is intrigung from the present point of view [14b]. How tunnelling into the negative-$U$ centres might actuallly occur is likewise intriguing, and it would be very valuable to see how the results are modified by employing a superconducting tip rather than the customary non-superconducting one – typically Au, W or Ir. Perhaps one can use drawn, stranded $Nb_3Sn$.

Already in [10a] I have pointed to an appropriate cluster geometry with which now one might try to press forward to a formal dynamic cluster analysis [80] of the relevant inhomogeneous *negative*-$U$ circumstance. The cluster suggested is quite large, but its analysis should become tractable within the near future. In order to encourage those whose *forte* this is to expend the required time and money, I would, in closing, like to draw attention to one final empirical result of great import for the current matter. An extensive study very recently has been released of a new system claimed to manifest strong evidence for superconductivity at 84 K [81]. The material is not a layered perovskite but a cubic perovskite related one, and it is, moreover, predominantly not a copper based system but a ruthenium based one. And yet the new material does contain copper. Indeed, from its highly oxidized stoichiometry, it must again contain Cu(III), for most certainly it cannot contain solely Cu(II) in the doping-induced presence of Ru(VI). Accordingly with this new find, Ba$_2$Y(Ru$_{1-u}$Cu$_u$)O$_6$ with $u \sim \frac{1}{6}$, one is back with the situation originally to confront us in LSCO, etc.. Namely of $^8$Cu$_{III}^0$ sites incorporated into a highly correlated $\sigma^*$ $d$-band, open to fluctuational conversion to $^{10}$Cu$_{III}^{2-}$, and with the latter resonantly attuned to $E_F$, as set by the one-electron system.

**Acknowledgement**

The author would like to thank the Leverhulme Foundation for invaluable financial support over the past three years through their Senior Fellowship scheme. My thanks go too to N.E. Hussey for continued stimulating discussions on HTSC matters.